\documentclass[useAMS,usenatbib]{mnras}
\usepackage{epsfig}
\usepackage{amsmath, amssymb,bm}
\usepackage[varg]{txfonts}
\usepackage{color}
\usepackage{aas_macros}

\title[Water delivery to the Earth from asteroids]{How much water was delivered from the asteroid belt to the Earth after its formation?}
\author[R. G. Martin et al.]{Rebecca G. Martin\thanks{E-mail:
    rebecca.martin@unlv.edu} and Mario Livio
\\Department of Physics and Astronomy, University of Nevada, Las
Vegas, 4505 South Maryland Parkway, Las Vegas, NV 89154, USA \\ }

\date{}

\pagerange{\pageref{firstpage}--\pageref{lastpage}} 
\pubyear{2020}

\begin{document}
\maketitle
\label{firstpage}
\begin{abstract} 
The Earth contains between one and ten oceans of water, including water within the mantle, where one ocean is the mass of water on the Earth's surface today. With $n$-body simulations we consider how much water could have been delivered from the asteroid belt to the Earth after its formation. Asteroids are delivered from unstable regions near resonances with the giant planets. We compare the relative impact efficiencies from the $\nu_6$ resonance, the 2:1 mean motion resonance with Jupiter and the outer asteroid belt. The $\nu_6$ resonance provides the largest supply of asteroids to the Earth, with about $2\%$ of asteroids from that region colliding with the Earth. Asteroids located in  mean motion resonances with Jupiter and in the outer asteroid belt have negligible Earth-collision probabilities. The maximum number of Earth collisions occurs if the asteroids in the primordial asteroid belt are  first moved into the $\nu_6$ resonance location (through asteroid-asteroid interactions or otherwise) before their eccentricity is excited sufficiently for  Earth collision. A maximum of about eight oceans of water may be delivered to the Earth. Thus, if the Earth contains ten or more oceans of water, the Earth likely formed with a significant fraction of this water. 
\end{abstract} 
  
 
\section{Introduction}   

The planets in the inner solar system are relatively dry \citep[e.g.][]{Abeetal2000}. The precise amount of water in and on the Earth is unknown, but is thought to be between one and ten "oceans" \citep[e.g.][]{Abeetal2000,Hirschmann2006,Mottl2007,Marty2012,Halliday2013},  where one ocean is about $2.5\times 10^{-4}\,\rm M_\oplus$ or the mass of water on the Earth's surface. Carbonaceous chrondrite (C-type) asteroids in the asteroid belt have a similar D/H ratio as the water on Earth, so their collisions with the young Earth are thought to have formed at least part of the oceans \citep[e.g.][]{Meech2019}. Earth's magma ocean lasted up to $100\,\rm Myr$ \citep{ElkinsTanton2012} and the Earth's oceans likely formed within 150-250$\,\rm Myr$ of the start of the solar system suggesting that the Earth's water arrived early \citep{Wilde2001,Mojzsis2001}.

The snow line is the radial location from the Sun outside of which water is found in a solid form. The snow line moves inwards during the evolution of the gas disc \citep{GaraudandLin2007,MartinandLivio2012,Martin2013}.  Inside of the snow line, water is in a gaseous form and may not be incorporated into the building blocks of planets, the planetesimals. The inner planets form with little water. However, there have been several suggestions of how the Earth could have formed with its current amount of water \citep[e.g.][]{Raymond2020}. The first possible way is through adsorption of hydrogen molecules on to silicate grains. 
Recently, \cite{Piani2020} suggested that the Earth's water could have formed from the hydrogen in  EC meteorites that formed the Earth. EC meteorites have a similar isotopic composition to terrestrial rocks. These meteorites are still in high abundance in the inner parts of the asteroid belt.  They suggest that these meteorites could have provided the Earth with up to three oceans worth of water. Second, it has been suggested that the Earth initially accreted a H-rich envelope that reacted with the Earth's surface magma ocean and generated water by hydrating silicates \citep{Ikoma2006}. Third, water may have been delivered by icy pebbles {\it if} the snowline passed inside the planet's orbit \citep{Sato2016,Ida2019}. Fourth, the terrestrial planets could have formed from material in a radially wide feeding zone \citep{Raymond2004,Raymond2009}. The late stages of terrestrial planet formation are thought to occur after the giant planets have formed \citep[e.g.][]{Wetherill1991,Chambers2001,Raymond2006,Fischer2014,Izidoro2015} and, in this scenario, the terrestrial planets continue to grow through collisional growth for up to $100\,\rm Myr$. The material which forms the terrestrial planets then may incorporate a small amount of material from the outer asteroid belt\footnote{While the classical model is unable to explain the low mass of Mars \citep[e.g.][]{Raymond2009} there are other viable explanations such as early giant planet instability  \citep[e.g.][]{Tsiganis2005,Morbidelli2007,Clement2018,Clement2019,Nesvorny2021}, the Grand Tack \citep{Walsh2011,OBrien2014,Jacobson2014} and inefficient planetesimal formation in the asteroid belt \citep{Levison2015,Izidoro2015,Raymond2017}.}.

The current leading scenario for the majority of the water delivery is by external pollution from the outer parts of the asteroid belt  as opposed to  local accretion \citep{Raymond2020}. It may have been from a few planetary embryos in the outer part of the asteroid belt \citep{Morbidellietal2000}  although the problem with stochastic delivery via large embryos in the outer belt is that, if such objects did form in the outer belt, they all need to be removed to prevent fossilizing gaps in the belt \citep{OBRien2011}. Asteroids and comets have certainly hit the Earth and delivered elements to its surface. Our Moon provides a record of the collisions \citep[e.g.][]{Tera1974}. The rate of collisions declined over  Gyr timescales as the belt is eroded \citep[e.g.][]{Minton2010}. The early collisions, at least, had the same size distribution as seen in the asteroid belt today \citep[e.g.][]{Strom2005,Bottke2005,Strom2015}.  Constraining asteroid belt delivery with crater counts is complicated as the majority of these impacts are left-over planetesimals \citep[e.g.][]{Morbidelli2018}. However, no matter how much water the Earth formed with, at least some water must have been delivered to the surface through later collisions. 

In this letter we estimate how much water could have been brought to the Earth through asteroid collisions after the formation of the Earth.  
%and the other planets in the solar system. 
 Asteroids that are located in resonance locations within the asteroid belt have their eccentricity increased until they are able to collide with the Earth \citep[e.g.][]{Bottke2002,Chen2019}. In section~\ref{model} we describe our $n$-body simulations in which we model three radially narrow regions of the asteroid belt: in the $\nu_6$ resonance, in the 2:1 mean motion resonance with Jupiter and in the chaotic region outside of the asteroid belt.  We compare the relative impact efficiencies between asteroids in these regions and the Earth. In section~\ref{discussion} we discuss implications for the delivery of water to the Earth and we conclude in section~\ref{concs}.

\section{Model of asteroid collisions}
\label{model}

\begin{table*}
\begin{tabular}{l l l l l l l l l l l l}
\hline
 Simulation    & $a_{\rm in}/{\rm au}$ & $a_{\rm out}/{\rm au}$ & $R_{\rm Earth}/R_\oplus$ & $N_{\rm Sun}$ & $N_{\rm Earth}$ & $N_{\rm J,S}$ &  $N_{\rm eject}$ &  $N_{\rm outcome}$ & $N_{\rm remain}$ & $P_{\rm collide}$   \\
 \hline
 \hline
run1 & 2.0   & 2.1  & 1  & 5269 & 113 & 1  & 524  & 5907 & 4093  &  0.019  \\
run1b & 2.0   & 2.1  & 10  & 4116 & 2730 & 0 & 254 & 7100 & 2900 &  0.38    \\ 
\hline 
run2 & 3.3   & 3.35 & 1  & 523 & 0  & 1  & 2658 & 3182 &6818  & -   \\
run2b & 3.3   & 3.35 & 10  & 510 & 18 & 1 & 2656  & 3185 &  6815 & 0.0057  \\
\hline
run3 & 4.0   & 4.1  & 1  & 27  & 0  & 34   & 8411& 8472 & 1528 &  - \\
run3b & 4.0   & 4.1  & 10  & 27 &  7 & 35 & 8419  & 8488 & 1512 & 0.00082\\ 
\hline
\end{tabular}
\caption{Outcomes of the simulations. Column~1 is the name of the simulation. Columns~2 and~3 are the inner and outer radius of the belt in the simulation, respectively. Column~4 shows the radius of the Earth. Columns~5 and~6 show the number of asteroids that hit the Sun and the Earth, respectively. Column~7 shows the number of asteroids that hit Jupiter or Saturn. Column~8 shows the number of asteroids that have been ejected.  Column~9 shows the number of asteroids that have an outcome (either collision or ejection). Column~10 shows the number of asteroids remaining in the simulation. Column~11 shows the probability of an Earth collision for all of the asteroids that have an outcome and are not still remaining in the simulation, $P_{\rm collide}=N_{\rm Earth}/N_{\rm outcome}$.  }
\label{table}
\end{table*}

\begin{figure*}
\hspace{-2.9cm}  
\includegraphics[width=4.5cm]{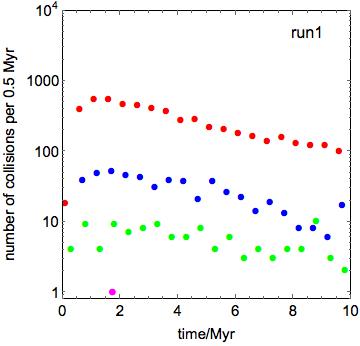} 
\includegraphics[width=4.5cm]{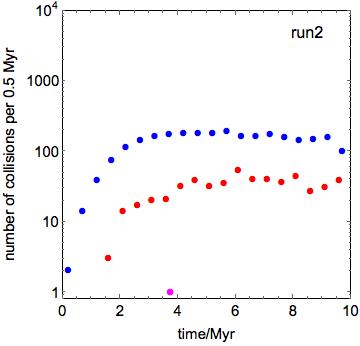}
\includegraphics[width=4.5cm]{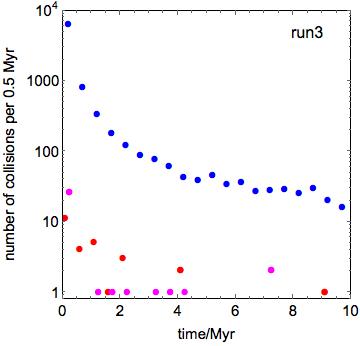}
\hspace{2.8cm} 
\includegraphics[width=4.5cm]{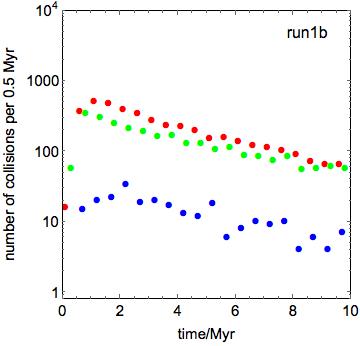}  
\includegraphics[width=4.5cm]{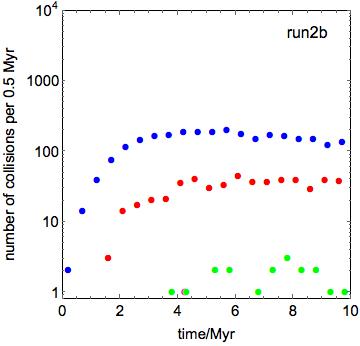} 
\includegraphics[width=7.3cm]{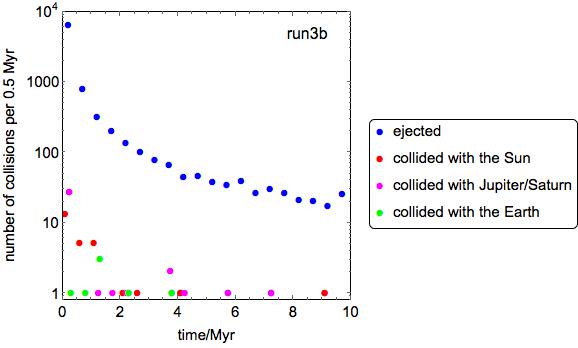} 
\caption{The outcomes of the $n$-body simulations in time binned into $0.5\,\rm Myr$ intervals. Note that the points are slightly offset from the centre of the time bin so that they don't completely overlap. 
The upper panels have the Earth radius of $R_{\rm Earth}=1\,\rm R_\oplus$  while the lower panels have Earth radius of $R_{\rm Earth}=10\,\rm R_\oplus$. The left panels show the simulation of the $\nu_6$ resonance in the range $a=2.0-2.1\,\rm au$. The middle panels show the 2:1 resonance in the range $a=3.3-3.35\,\rm  au$. The right panels show the simulation in the region $4-4.1\,\rm au$.   The blue points show asteroids that are ejected. The red points show asteroids that hit the Sun. The green points show asteroids that hit the Earth. The magenta points show asteroids that hit Jupiter or Saturn.} 
\label{main} 
\end{figure*}

We use the Bulirsch-Stoer integrator in the {\sc mercury} code \citep{Chambers1999} to model the evolution and fate of asteroids that begin in the asteroid belt.  We include the planets Jupiter, Saturn and Earth and a belt of asteroids in each simulation.  The fraction of asteroids that collide with Mars is a factor of at least a few smaller than the collisions with Earth \citep{Ito2006,Minton2010}. Although we note that since we do not include Mars, this may lead to a slightly increased collision rate with the Earth. The asteroids are treated as test particles that do not interact with each other. We assume that the planets are fully formed and we ignore gas friction  in our simulations. 

In order to get a statistically significant number of asteroid collisions we focused on  three radially narrow regions of the asteroid belt: (1) the $\nu_6$ resonance at about $2.1\,\rm au$ \citep[e.g.][]{Froeschle1986,Morbidelli1991,Morbidelli1994}  (this is the zero-inclination, eccentricity-averaged location), (2) the 2:1 mean motion resonance with Jupiter at about $3.3\,\rm au$ \citep[e.g.][]{Smallwood2018} and (3) the chaotic region outside of the asteroid belt. The simulation parameters are all described in Table~\ref{table}.  We are interested here in the fraction of Earth collisions and so we first ran simulations in which we did not inflate the size of the Earth (run1, run2 and run3).   Since we did not get collisions with the Earth in all cases, we  then  also considered three further simulations with an Earth radius of $R_{\rm Earth}=10\,\rm R_\oplus$ (run1b, run2b and run3b).  We analytically estimated then the probability of an Earth collision had we not inflated the size of the Earth  in order to estimate the fraction of Earth collisions in a belt with more particles than we are able to simulate. We considered a particle to be ejected if its semi-major axis became larger than $20\,\rm au$.  We ran each simulation for $10\,\rm Myr$. The Earth collision probability, $P_{\rm collide}$, is the probability of an Earth collision out of the asteroids that have a final outcome, where a final outcome is either a planet or stellar collision  or an  ejection. Since the timescale we ran our simulations for was longer than the resonance operation timescales \citep[e.g.][]{Bottke2002}, this relatively short timescale should not have affected our conclusions.

\subsection{The inner edge of the asteroid belt: the $\nu_6$ resonance} 
\label{nu6}

We first considered the fate of asteroids that begin in the $\nu_6$ resonance. We placed 10,000 test particles in the radial range $2-2.1\,\rm au$ distributed uniformly in radius. The eccentricities were uniformly distributed in the range $0-0.1$, and the inclinations were uniformly distributed in the range $0-10^\circ$.  The longitude of ascending node, the argument of perihelion and the mean anomaly were all uniformly distributed in the range $0-360^\circ$. The orbits of the planets were set to be the current orbits for each planet since the solar system planets have stable orbits over long time-scales \citep{Duncan1998,Ito2002}.  We note that the eccentricity and inclination distributions represent a dynamically cold and massive planetesimal disc that is applicable shortly after the gas disc has dissipated. These assumptions are skewed in order to estimate the upper limit of delayed delivery from the belt to the Earth.\footnote{The belt would be calm after the Earth’s formation in a late giant planet instability scenario \citep{Roig2015,Deienno2016} with a  primordially low-mass \citep[e.g.][]{Levison2015,Izidoro2015}.  However, the giant planet orbits would not be on their current day orbits.  The early giant planet instability models predict a much reduced asteroid belt at this time \citep[e.g.][]{Deienno2018,Clement2019}.  Chaotic excitation \citep{Izidoro2016} or self-stirring from embryos \citep{Petit2001} would have occurred early as well.  \cite{Minton2010} showed that the belt has been eroded by about $50\%$ over the past $4\,\rm Gyr$ of static evolution. Thus, the belt may already have been eroded by a factor of a few by the time the Earth and the giant planets were formed on their current orbits.}

The outcomes for run1 are shown in the top left panel of Fig.~\ref{main}. There are 113 Earth collisions during the simulation time of $1\times 10^7 \,\rm yr$. There are 4093 particles remaining in the simulation at the end. Thus, the number of Earth collisions compared to the number of particles that have an outcome already determined is $P_{\rm collide}=113/5907 =0.019 $. Therefore the $\nu_6$ resonance is very efficient at delivering asteroids to the Earth.   \cite{Ito2006} also considered the collision probability of asteroids with the Earth from the $\nu_6$ resonance and found a slightly higher probability of about 3\%. The difference is likely a result of the higher initial inclinations used in our simulations. Note that \cite{Morbidelli1998} and \cite{Gladman1997} considered similar calculations but with a smaller number of particles.

For each simulation we considered, we ran two different Earth sizes: one with the Earth at its current size, and one with an inflated Earth size of $R_{\rm Earth}=10\,\rm R_\oplus$. This is because some of our simulations with an uninflated Earth did not have any Earth collisions. The effective cross section of the Earth is given by
\begin{equation}
    A(R_{\rm Earth})=\pi R_{\rm Earth}^2 \left( 1+\frac{4 G M_\oplus}{R_{\rm Earth} v_{\rm rel}^2}  \right), 
\end{equation}
where $v_{\rm rel}$ is the relative velocity between the Earth and the asteroid.
If $v_{\rm rel}$ is high compared to the planet escape velocity, the number of collisions would increase by a factor of 100. However, if $v_{\rm rel}$ is slow compared to the escape velocity then the factor could be as low as 10, as a result of gravitational focusing.  The lower left panel of Fig.~\ref{main} shows run1b which has exactly the same initial conditions as run1 but the Earth is inflated up to a radius of $R_{\rm Earth}=10\,\rm R_\oplus$. As shown in Table~\ref{table}, the increase in the planet size led to the number of Earth collisions increasing by a factor of 20. 

We can estimate the factor by which the effective cross section changes with the increase in planet radius as $f=A(10{\,\rm R_\oplus})/A({\rm R_\oplus})$. A particle with semi-major axis of $2\,\rm au$ and eccentricity $e=0.5$ comes close to the Earth. If the orbit of the particle is in the same plane as the Earth, we can find the relative velocity as the difference in the periastron speed of the asteroid and the speed of the Earth. The relative increase in the effective cross section is $f=23$ and this is similar to that found in the simulations. If the mechanism works in the same way for particles coming from other resonances, then we can estimate how the cross section affects the number of collisions in those cases. For a particle at $3.3\,\rm au$ with eccentricity $e=0.69$ we find $f=31$. For a particle at $4\,\rm au$ with eccentricity $0.75$ we find $f=33$. We use these for our nominal values to calculate an approximate collision rate without the inflated Earth in the following sections.

\subsection{The 2:1 mean motion resonance with Jupiter}

The top middle panel of Fig.~\ref{main} shows run2 in which we model a portion of the 2:1 resonance with Jupiter between $3.3-3.35\,\rm au$. The other orbital parameters are distributed as described in the previous section. In this simulation there are no collisions with the Earth in the simulation up to a time of $10\,\rm Myr$. The lower middle panel of Fig.~\ref{main} shows run2b which represents the same initial setup except that the size of the Earth has been inflated to $10\,\rm R_\oplus$. In this case we find 18 collisions with the Earth. This gives a collision probability of $P_{\rm collide}=18/3185=0.0057$.  To account for the inflated Earth, this probability must be reduced. As we discussed in the previous section, for the 2:1 resonance location we expect this to be a factor of about 31. This gives a collision probability of $0.02\%$. Given that 3182 particles have had an outcome, the expected number of collisions is $<1$ and it is not surprising that we did not find any collisions in our simulation run2. The 2:1 resonance is not an efficient way to deliver material to the Earth since most asteroids are instead ejected from the inner solar system as a result of close encounters with Jupiter or Saturn. 

We note that the impact rate with the Earth from the 2:1 resonance may be slightly higher if higher inclination asteroids were included. Overlapping secular and mean-motion resonances can help to move asteroids on to near Earth orbits more efficiently. In the outer belt, the $\nu_5$ resonance is at about $i=25^\circ$ for $e=0.1$ and the $\nu_6$ resonance is at about $i=20^\circ$ \citep{Morbidelli1991}. However, even with some higher inclination asteroids included, the rate of Earth collisions is unlikely to have been as significant as from the $\nu_6$ resonance at $2\,\rm au$ with low inclination asteroids.

\subsection{The outer edge of the asteroid belt}

Finally, we considered a radially narrow region in the chaotic region outside of the current location of the asteroid belt at $4.0-4.1\,\rm au$. The right hand panels in Fig.~\ref{main} show the outcomes for the particles. With the uninflated Earth (top) there were no collisions. With the inflated Earth (bottom) there were 7 collisions with a collision probability of $7/8488=0.08\%$. As discussed in section~\ref{nu6}, we must divide this by 33 to get the expected probability without the inflated Earth and we find a probability of about $0.0025\%$. Thus, the chaotic region itself it not an efficient way to get asteroids to the Earth.

\section{Delivery of water from the asteroid belt}
\label{discussion}

The current mass of water in the Earth's oceans is $M_{\rm ocean}=2.5\times 10^{-4}\,\rm M_\oplus$. The current mass of the asteroid belt is about $4.5\times 10^{-4}\,\rm M_\oplus$ \citep[e.g.][]{Krasinsky2002,Kuchynka2013,DeMeo2013}. The original mass of the asteroid belt was up to a few thousand times more massive \citep[e.g.][]{Morbidellietal2000,Petit2001}. Following recent terrestrial planet formation simulations, we take an upper limit to the initial asteroid belt mass of $M_{\rm belt, i}=2\,\rm M_\oplus$ \citep[e.g.][]{OBrien2007,Deienno2018,Clement2019}.  This is calculated by interpolating the solid material contained in the planets \citep{Weidenschilling1977,Hayashi1981}.

The composition of the asteroid belt varies with orbital radius \citep[e.g.][]{Gradie1982,DeMeo2014}. 
While the asteroids today are not completely separated by type at the snow line, there is a clear jump in the composition at the snow line radius, $R_{\rm snow}=2.7\,\rm au$. The inner parts are dominated by S-type asteroids that contain a fraction of water up to $f_{\rm water,S}=0.001$. C-type asteroids dominate the outer parts of the asteroid belt. These contain a fraction of water of about $f_{\rm water,C }=0.1$ \citep[e.g.][]{Kerridge1985,Alexander2018}. We estimate the initial relative fraction of C-type asteroids by assuming a surface density profile $\Sigma \propto R^{-3/2}$ between $R_{\rm in}=1.55\,\rm au$ (three Hill radii from Mars) up to $R_{\rm out}=4.1\,\rm au$ (three Hill radii from Jupiter). The fraction of mass contained in $R_{\rm snow}<R<R_{\rm out}$ compared to the whole belt is $f_{\rm C}=0.49$.
The total amount of water there initially was about $0.1\,{\rm M_\oplus}\approx 400\,M_{\rm ocean}$. Thus there is a large supply of water in these C-type asteroids that could have been delivered to Earth. The majority of the water that came from the asteroid belt must have come from C-type asteroids.  

In section~\ref{model} we showed that that the collision probability for asteroids with the Earth is much higher from the $\nu_6$ resonance than any mean motion resonances farther out or the chaotic region. Asteroids do not necessarily have to form close to the $\nu_6$ to be delivered to the Earth from there.  
%Resonances are not a finite reservoir of objects that goes away once depleted.  
There are other mechanisms such as gas drag, asteroid-asteroid interactions and the Yarkovsky effect \citep[e.g.][]{Farinella1998}  which may move asteroids into this resonance region.  \cite{Morbidelli1999} found that the numerous weak resonances in the asteroid belt allow asteroids to migrate in eccentricity and move through the belt. Furthermore, \cite{Raymond2017} suggest that asteroids may be implanted into the asteroid belt region from farther out. Observational evidence also suggests that this is possible \citep[e.g.][]{Jewitt2014}.
%Volatile-rich C-type asteroids still make up around half of the asteroids in the inner belt, thus this material could deliver a lot of water to the Earth.  

We now consider the maximum amount of water that could have been delivered to Earth if we assume that the asteroids from the outer parts of the belt can move into the $\nu_6$ resonance. Considering only the C-type asteroids, we find the amount of water delivered (wd), assuming complete efficiency, to be 
\begin{equation}
\frac{M_{\rm wd}}{M_{\rm ocean}}\approx 8 
\left(\frac{P_{\nu_6,\rm collide,}}{0.02}\right) 
\left(\frac{f_{\rm C}}{0.49}\right) 
\left(\frac{f_{\rm water,C}}{0.1}\right)
\left(\frac{M_{\rm belt, i}}{ 2\,\rm M_\oplus}\right).
\end{equation}
This mechanism therefore may have been able to deliver the Earth's water as it is today, but it is unlikely to have delivered the upper limit of 10 oceans that may be contained within the mantle of the Earth. If there is indeed such a large reservoir of water in the Earth, it seems likely that the Earth formed with it. 

The collision probabilities from the 2:1 resonance and the chaotic region are significantly smaller, it is unlikely that either region could have supplied enough material, unless a particularly large water-rich asteroid happened to be in the region.
We have used a dynamically cold belt (low eccentricity and low inclination orbits). A dynamically warmer belt may have a slightly increased collision rate with the Earth from the 2:1 resonance due to the overlapping $\nu_5$ and $\nu_6$ resonances at higher inclinations.

\section{Conclusions}
\label{concs}

 With $n$-body simulations we have compared the relative impact efficiencies from the $\nu_6$ resonance, the 2:1 mean motion resonance with Jupiter and the outer asteroid belt after the planets have formed. 
Our analysis assumed that the asteroid belt contained its primordial mass after the Earth had formed and the giant planets were on their current day orbits. Thus, we have skewed the assumptions to estimate an upper limit to the amount of water that could have been delivered to the Earth. We have shown that the majority of asteroid collisions with the Earth originate from the $\nu_6$ resonance at the inner edge of the the asteroid belt. About $2\%$ of asteroids from the resonance collide with the Earth. The collision probability from the 2:1 mean motion resonance is about one hundred times smaller and from the chaotic region about a thousand times smaller. We estimate that if the majority of asteroids in the primordial asteroid belt were moved into the $\nu_6$ resonance either through asteroid-asteroid interactions or gas drag, or the Yarkovsky effect, then at most, the asteroid belt could have delivered about eight oceans worth of water.  Thus, the delivery of one ocean's worth from the asteroid belt was certainly possible. However, the delivery of 10 oceans worth could have been difficult and if the Earth's mantle contains such significant amounts of water then the Earth likely formed with a good fraction of it, as recently suggested by \cite{Piani2020}.

\section*{Data availability statement}
The results in this paper can be reproduced using the {\sc mercury} code (Astrophysics Source Code Library identifier {\tt ascl.net/1201.008}). The data underlying this article will be shared on reasonable request to the corresponding author.

\section*{Acknowledgements} 
RGM acknowledges support from NASA through grant 80NSSC21K0395. Computer support was provided by UNLV's National Supercomputing Center. 
 
\bibliographystyle{mnras}
\bibliography{ms} 
\label{lastpage} 
\end{document}